\documentclass[
  aip,
  amsmath,amssymb,
  preprint
]{revtex4-1}

\usepackage{graphicx}
\usepackage{amsmath,amssymb,amsthm}
\usepackage{booktabs}
\usepackage{array}
\usepackage{enumitem}
\usepackage{tikz}
\usetikzlibrary{patterns,calc,arrows.meta}
\usepackage{algpseudocode}
\usepackage{url} % optional but safe

\usepackage{hyperref}

\newtheorem{theorem}{Theorem}[section]

\newtheorem{proposition}[theorem]{Proposition}

\newtheorem{remark}{Remark}[section]

\begin{document}
\raggedbottom

\title{The Zero-Frequency Limit of Spherical Cavity Modes: On the Formal Endpoint at $\nu = -1$}

\author{Mustafa Bakr}
\email{mustafa.bakr@physics.ox.ac.uk}
\affiliation{Clarendon Laboratory, Department of Physics, University of Oxford}
\author{Smain Amari}
%\date{\today}

\begin{abstract}
The transverse magnetic (TM) modes of a spherical cavity satisfy a dispersion  relation connecting the angular eigenvalue $\nu$ to the resonant frequency through zeros of the spherical Bessel function derivative. Analytic continuation of this dispersion relation to $\nu = -1$ yields a formal zero-frequency endpoint where $j_{-1}(x) = \cos x / x$ admits the root $x = 0$. We examine this limit in detail, showing that while the mathematics is well-defined, the endpoint does not correspond to a physical electromagnetic mode. The positivity of the angular Sturm-Liouville operator restricts physical eigenvalues to $\nu \geq 0$, placing $\nu = -1$ outside the admissible spectrum. We demonstrate that all electromagnetic field components vanish in this limit, even though the underlying Debye potential 
$\Pi = \cos(kr)/kr$ remains non-trivial and exhibits a monopole-type singularity at the origin. This distinction between potential and field reflects the kernel structure of the curl-curl operator for spherically symmetric configurations. The analysis clarifies the boundary between propagating electromagnetic modes and static field configurations in spherical geometry, connecting the formal endpoint to longstanding questions about mode counting in cavity quantization.
\end{abstract}

\maketitle

\section{Introduction}

The electromagnetic modes of a perfectly conducting spherical cavity of radius $a$ are characterised by discrete resonant frequencies determined by boundary conditions at the cavity wall. For transverse magnetic (TM) modes, the resonant condition requires
\begin{equation}
\left.\frac{d}{dx}\left[x j_\nu(x)\right]\right|_{x=ka} = 0,
\label{eq:TM_BC}
\end{equation}
where $j_\nu(x)$ is the spherical Bessel function of order $\nu$, $k = \omega/c$ is the wavenumber, and $\nu$ is the angular eigenvalue arising from separation of variables in the Helmholtz equation~\cite{Stratton1941,Jackson1999}. On the full sphere with regularity demanded at both poles, $\nu$ is restricted to non-negative integers $\ell = 0, 1, 2, \ldots$, yielding the familiar quantised spectrum.

Recent work has shown that the angular eigenvalue $\nu$ may be extended continuously beyond integer values when the spherical domain is modified by conducting cones or wedges~\cite{bakrsphere, BakrAmari2025, BakrAmari2023, BakrAmari2025, bakr2025quantummechanicssphericalwedge}. This raises a natural question: what are the boundaries of the continuous spectrum? The TM dispersion relation~\eqref{eq:TM_BC} admits a formal zero-frequency solution when $\nu = -1$, since
\begin{equation}
j_{-1}(x) = \frac{\cos x}{x}
\label{eq:j_minus1}
\end{equation}
and the derivative $[x j_{-1}(x)]' = -\sin x$ vanishes at $x = 0$. This suggests that $\nu = -1$ represents a zero-frequency endpoint of the TM dispersion curve.

The purpose of this paper is to examine this formal endpoint carefully. We demonstrate that while the mathematical structure is well-defined, the limit $\nu \to -1$ does not correspond to a physical electromagnetic mode. The angular Sturm-Liouville operator possesses a positivity constraint that excludes the region $-1 < \nu < 0$ from the physical spectrum, and field components vanish identically as this limit is approached. The analysis clarifies the mathematical boundary of the mode spectrum and its connection to electrostatic configurations.

This analysis complements the treatment of the $(\nu, m) = (0, 0)$ boundary point in Ref.~\cite{BakrAmari2025}, where the same eigenvalue 
$\lambda = \nu(\nu+1) = 0$ is approached from $\nu = 0$ rather than $\nu = -1$. Both limits yield vanishing electromagnetic fields but non-trivial potentials, reflecting the two roots of the quadratic eigenvalue relation.

\section{The Angular Eigenvalue Problem}

\subsection{Sturm-Liouville Structure}

The angular dependence of electromagnetic modes in spherical coordinates is governed by the associated Legendre equation
\begin{equation}
\frac{1}{\sin\theta}\frac{d}{d\theta}\left(\sin\theta\frac{d\Theta}{d\theta}\right) + \left[\lambda - \frac{m^2}{\sin^2\theta}\right]\Theta = 0,
\label{eq:angular_ODE}
\end{equation}
where $\lambda = \nu(\nu+1)$ is the separation constant and $m$ is the azimuthal index. This equation defines a singular Sturm-Liouville problem on the interval $\theta \in (0, \pi)$ with regular singular points at both poles.

The Sturm-Liouville operator
\begin{equation}
\mathcal{L} = -\frac{1}{\sin\theta}\frac{d}{d\theta}\left(\sin\theta\frac{d}{d\theta}\right) + \frac{m^2}{\sin^2\theta}
\label{eq:SL_operator}
\end{equation}
acts on the weighted Hilbert space $L^2((0,\pi), \sin\theta\, d\theta)$. With appropriate boundary conditions ensuring regularity at both poles, this operator is self-adjoint and positive semi-definite~\cite{Zettl2005}. Its eigenvalues satisfy
\begin{equation}
\lambda = \nu(\nu+1) \geq 0.
\label{eq:positivity}
\end{equation}

\subsection{The Forbidden Region}

The positivity constraint~\eqref{eq:positivity} has immediate consequences for the admissible values of $\nu$. The equation $\nu(\nu+1) = \lambda$ has solutions
\begin{equation}
\nu = -\frac{1}{2} \pm \sqrt{\frac{1}{4} + \lambda}.
\label{eq:nu_solutions}
\end{equation}
For $\lambda \geq 0$, the two roots are
\begin{equation}
\nu_+ = -\frac{1}{2} + \sqrt{\frac{1}{4} + \lambda} \geq 0, \qquad \nu_- = -\frac{1}{2} - \sqrt{\frac{1}{4} + \lambda} \leq -1.
\label{eq:nu_branches}
\end{equation}
The conventional choice $\nu = \nu_+ \geq 0$ ensures that the spherical Bessel function $j_\nu(kr)$ remains regular at the origin.

The region $-1 < \nu < 0$ corresponds to $\lambda < 0$, which violates the positivity of the Sturm-Liouville operator. No physical eigenmode can have an angular eigenvalue in this forbidden interval. The boundary points $\nu = 0$ and $\nu = -1$ both correspond to $\lambda = 0$, representing the edge of the physical spectrum.

\begin{proposition}[Forbidden Region]
\label{prop:forbidden}
For the angular Sturm-Liouville problem~\eqref{eq:angular_ODE} with regularity at both poles, no eigenvalue $\nu$ satisfies $-1 < \nu < 0$. The physical spectrum is restricted to $\nu \geq 0$.
\end{proposition}

\section{The Limit $\nu \to -1$}

\subsection{Radial Functions}

The spherical Bessel function $j_\nu(x)$ admits analytic continuation to non-integer and negative orders through the relation
\begin{equation}
j_\nu(x) = \sqrt{\frac{\pi}{2x}} J_{\nu+1/2}(x),
\label{eq:spherical_bessel}
\end{equation}
where $J_{\nu+1/2}$ is the ordinary Bessel function. For $\nu = -1$, this yields
\begin{equation}
j_{-1}(x) = \sqrt{\frac{\pi}{2x}} J_{-1/2}(x) = \sqrt{\frac{\pi}{2x}} \cdot \sqrt{\frac{2}{\pi x}} \cos x = \frac{\cos x}{x}.
\label{eq:j_minus1_derivation}
\end{equation}
The function $j_{-1}(x)$ diverges as $x \to 0$, in contrast to $j_\nu(x)$ for $\nu > -1$ which vanishes as $x^\nu$ at the origin.

The TM boundary condition involves
\begin{equation}
\frac{d}{dx}\left[x j_{-1}(x)\right] = \frac{d}{dx}[\cos x] = -\sin x,
\label{eq:TM_derivative}
\end{equation}
which vanishes at $x = 0, \pi, 2\pi, \ldots$ The root $x = ka = 0$ formally corresponds to zero frequency, $\omega = 0$.

\subsection{Angular Functions}

The Legendre function $P_\nu(\cos\theta)$ at $\nu = -1$ takes a particularly simple form. Using the integral representation or hypergeometric series, one finds~\cite{DLMF}
\begin{equation}
P_{-1}(\cos\theta) = 1
\label{eq:P_minus1}
\end{equation}
for all $\theta \in [0, \pi]$. This can be verified by direct substitution into the Legendre equation with $\lambda = \nu(\nu+1) = (-1)(0) = 0$ and $m = 0$:
\begin{equation}
\frac{1}{\sin\theta}\frac{d}{d\theta}\left(\sin\theta\frac{d(1)}{d\theta}\right) + 0 \cdot 1 = 0. \quad 
\label{eq:P_minus1_verification}
\end{equation}
The constant function $\Theta = 1$ is indeed a solution when $\lambda = 0$.

\subsection{Field Components}

To determine whether the formal solution at $\nu = -1$ corresponds to a physical electromagnetic mode, we examine the field components derived from the Debye potential formalism. For TM modes, the fields are~\cite{Stratton1941}
\begin{align}
E_r &= \frac{\nu(\nu+1)}{r^2} j_\nu(kr) P_\nu(\cos\theta), \label{eq:Er}\\
E_\theta &= \frac{1}{r}\frac{d}{dr}[r j_\nu(kr)] \frac{dP_\nu}{d\theta}, \label{eq:Etheta}\\
H_\phi &= -i\omega\varepsilon\, j_\nu(kr) \frac{dP_\nu}{d\theta}. \label{eq:Hphi}
\end{align}

We now evaluate each component in the limit $\nu \to -1$:

\paragraph{Radial electric field.} From Eq.~\eqref{eq:Er},
\begin{equation}
E_r = \frac{\nu(\nu+1)}{r^2} j_\nu(kr) P_\nu(\cos\theta) \xrightarrow{\nu \to -1} \frac{(-1)(0)}{r^2} j_{-1}(kr) \cdot 1 = 0.
\label{eq:Er_limit}
\end{equation}
The factor $\nu(\nu+1) \to 0$ annihilates $E_r$ regardless of the radial function's behaviour.

\paragraph{Polar electric field.} From Eq.~\eqref{eq:Etheta},
\begin{equation}
E_\theta = \frac{1}{r}\frac{d}{dr}[r j_\nu(kr)] \frac{dP_\nu}{d\theta} \xrightarrow{\nu \to -1} \frac{1}{r}\frac{d}{dr}[r j_{-1}(kr)] \cdot \frac{d(1)}{d\theta} = 0.
\label{eq:Etheta_limit}
\end{equation}
The angular derivative $dP_{-1}/d\theta = 0$ annihilates $E_\theta$.

\paragraph{Azimuthal magnetic field.} From Eq.~\eqref{eq:Hphi},
\begin{equation}
H_\phi = -i\omega\varepsilon\, j_\nu(kr) \frac{dP_\nu}{d\theta} \xrightarrow{\nu \to -1} -i\omega\varepsilon\, j_{-1}(kr) \cdot 0 = 0.
\label{eq:Hphi_limit}
\end{equation}
Again, the vanishing angular derivative eliminates the field component.

The remaining components ($E_\phi$, $H_r$, $H_\theta$) vanish for TM modes with $m = 0$ by symmetry. Thus all electromagnetic field components vanish identically in the limit $\nu \to -1$.

\begin{theorem}[Vanishing Fields at $\nu = -1$]
\label{thm:vanishing}
The TM electromagnetic field components derived from a Debye potential with angular dependence $P_\nu(\cos\theta)$ satisfy
\begin{equation}
\lim_{\nu \to -1} \mathbf{E} = \mathbf{0}, \qquad \lim_{\nu \to -1} \mathbf{H} = \mathbf{0}.
\label{eq:vanishing_theorem}
\end{equation}
The formal solution at $\nu = -1$ generates no electromagnetic field.
\end{theorem}

\begin{remark}[Non-trivial potential at $\nu = -1$]
While all electromagnetic field components vanish as $\nu \to -1$, the 
Debye potential
\begin{equation}
\Pi = j_{-1}(kr) P_{-1}(\cos\theta) = \frac{\cos(kr)}{kr}
\end{equation}
remains non-trivial. Moreover, this potential is \emph{singular} at the 
origin:
\begin{equation}
\Pi \sim \frac{1}{kr} \quad (r \to 0).
\end{equation}
This singularity has the same local structure as the Green's-function 
potential of a point monopole and connects formally to the electrostatic 
configuration discussed in Section~\ref{sec:monopole}, although no physical 
source is present in the cavity problem. The vanishing of the electromagnetic 
field despite a singular potential reflects the structure of the curl-curl 
operator that extracts fields from the Debye potential: for spherically 
symmetric configurations, this operator has a non-trivial kernel.

This contrasts with the situation at $\nu = 0$, where the potential 
$\Pi = j_0(kr) = \sin(kr)/kr$ is regular at the origin ($\Pi(0) = 1$) 
yet still generates no electromagnetic field. Both roots of the 
eigenvalue equation $\nu(\nu+1) = 0$ yield vanishing fields but with 
qualitatively different potential structures: regular (source-free) at 
$\nu = 0$, and singular (monopole-type) at $\nu = -1$.

This distinction clarifies the role of endpoint solutions in mode counting 
and reinforces that the exclusion of $\nu = -1$ follows from physical 
admissibility rather than algebraic inconsistency.
\end{remark}

\section{Physical Interpretation}

\subsection{The Electrostatic Limit}

The vanishing of all field components as $\nu \to -1$ is not a mathematical accident but reflects a deep physical constraint. The separation constant $\lambda = \nu(\nu+1)$ appears in the angular equation as the eigenvalue of the angular Laplacian. When $\lambda = 0$, the angular equation reduces to
\begin{equation}
\nabla^2_{\Omega} \Theta = 0,
\label{eq:laplace_angular}
\end{equation}
whose only regular solution on the full sphere is the constant $\Theta = 1$. A constant angular dependence means no angular variation in the fields—but a time-harmonic electromagnetic field with no angular variation and finite energy in a bounded cavity cannot exist.

The situation differs for electrostatics. A point charge at the origin produces a purely radial electric field $\mathbf{E} = (Q/4\pi\varepsilon_0 r^2)\hat{r}$ with constant angular dependence. This configuration is excluded from the cavity mode spectrum by the requirement of finite field energy—the electrostatic field of a point charge has infinite energy within any volume containing the charge. The formal limit $\nu \to -1$ thus represents the boundary between propagating electromagnetic modes and singular electrostatic configurations.

\subsection{Connection to the Monopole}
\label{ec:monopole}
\begin{figure}[h]
\centering
\includegraphics[width=\columnwidth]{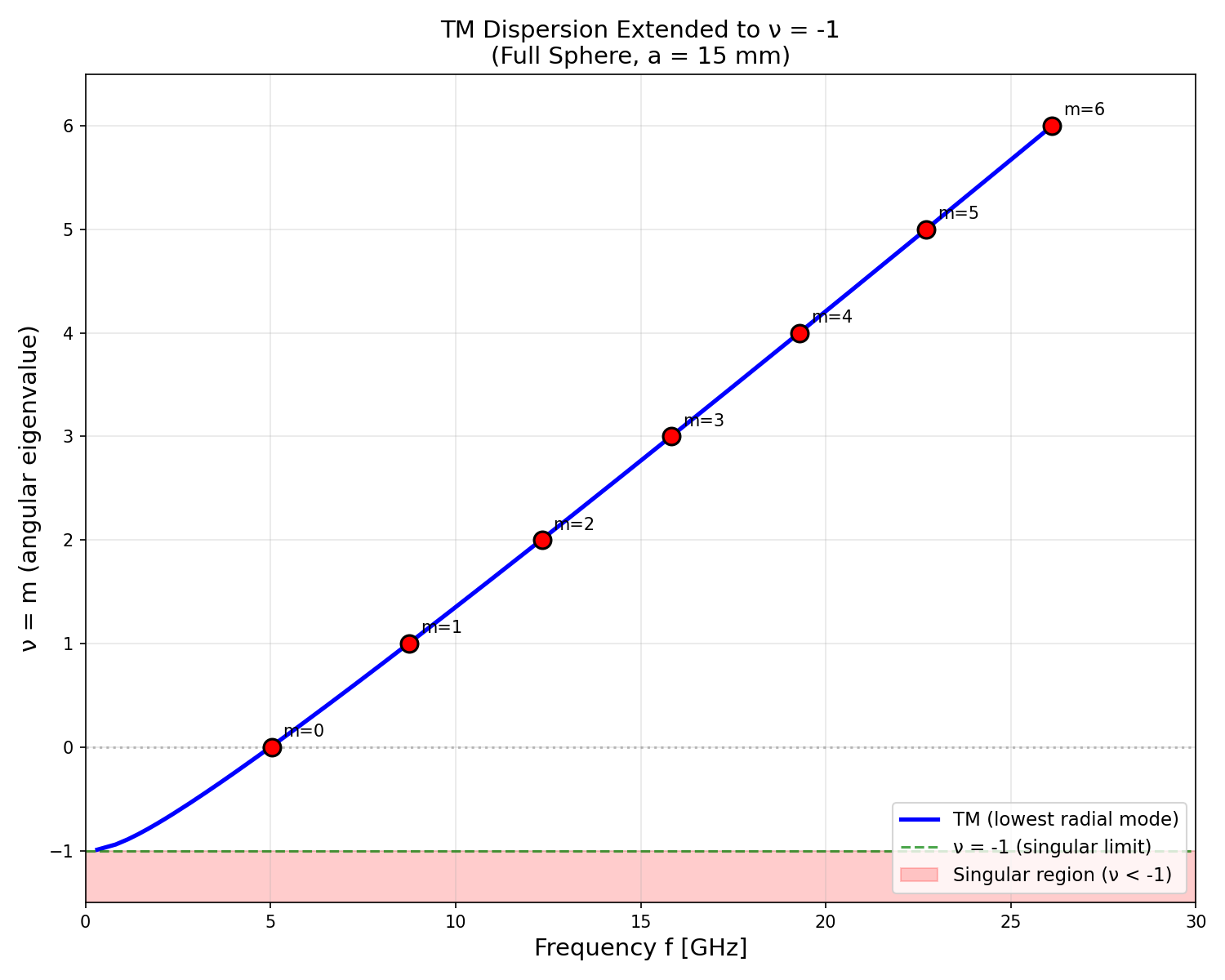}
\caption{TM dispersion relation for a spherical cavity, showing the first Bessel derivative root $x'_{\nu,1}$ as a function of angular eigenvalue $\nu$. The physical spectrum (solid curve, $\nu \geq 0$) connects integer modes (filled circles) through a continuous dispersion curve accessible via conical boundary modifications. The formal continuation to $\nu = -1$ (dashed curve) approaches $x = 0$ as $\nu \to -1^{-}$, corresponding to zero frequency. The shaded region $-1 < \nu < 0$ is forbidden by the positivity constraint $\nu(\nu+1) \geq 0$ of the angular Sturm-Liouville operator. The endpoint $\nu = -1$ represents a mathematical boundary where all electromagnetic field components vanish, connecting the cavity mode spectrum to electrostatic configurations.}
\label{fig:dispersion}
\end{figure} 
The angular function $P_{-1}(\cos\theta) = 1$ corresponds to spherical symmetry—no $\theta$-dependence whatsoever. In the context of Maxwell's equations, a spherically symmetric electric field $\mathbf{E} = E_r(r)\hat{r}$ satisfies $\nabla \cdot \mathbf{E} = 0$ only if $E_r \propto 1/r^2$, which is the Coulomb field that \emph{would} arise from a point charge. In the cavity problem, no such source exists; the singular potential structure represents a formal limit of the mode spectrum rather than a physical source configuration. The corresponding magnetic field configuration—a spherically symmetric $\mathbf{H} = H_r(r)\hat{r}$—would require a magnetic monopole, which does not exist in classical electrodynamics.

The formal structure at $\nu = -1$ thus touches upon the asymmetry between electric and magnetic sources in Maxwell's theory. The electric monopole (point charge) exists but produces static fields excluded from the cavity spectrum. The magnetic monopole does not exist, so no corresponding magnetic configuration arises. The endpoint $\nu = -1$ marks the boundary where the cavity mode formalism encounters these fundamental constraints.

\subsection{The Dispersion Curve Boundary}

Figure~\ref{fig:dispersion} illustrates the TM dispersion relation. The physical modes with $\nu = 0, 1, 2, \ldots$ lie on discrete points sampling a continuous curve. The curve extends smoothly toward $\nu = -1$ as frequency approaches zero, but the positivity constraint $\nu(\nu+1) \geq 0$ excludes the region $-1 < \nu < 0$ from the physical spectrum. The point $\nu = -1$ represents an asymptotic boundary that the physical spectrum approaches but cannot cross.
For cavities with conical truncations that access continuous values of $\nu$, the dispersion curve $\nu(\theta_c)$ remains bounded below. As the cone angle $\theta_c$ increases, $\nu$ decreases toward zero but cannot become negative. The minimum physical value $\nu = 0$ corresponds to the zonal mode $P_0(\cos\theta) = 1$, which for TM polarisation yields the lowest cavity resonance at $ka = \pi/2$.

\section{Relationship to TE Modes}

For completeness, we examine the corresponding limit for transverse electric (TE) modes. The TE boundary condition requires $j_\nu(ka) = 0$, whose first root for $\nu = 0$ is $x_{0,1} = \pi$, corresponding to a finite frequency. As $\nu \to -1$, the function $j_{-1}(x) = \cos x / x$ has zeros at $x = \pi/2, 3\pi/2, \ldots$, which remain at finite, nonzero values.

There is no zero-frequency limit for TE modes via the $\nu \to -1$ path. The asymmetry between TM and TE modes in this regard reflects their different physical character: TM modes have radial electric field components that can, in principle, connect to electrostatic configurations, while TE modes have no radial electric field and cannot approach electrostatic limits.

\section{Conclusion}

The formal endpoint $\nu = -1$ of the spherical cavity TM dispersion relation 
represents a mathematical boundary rather than a physical mode. Three independent 
constraints exclude this point from the physical spectrum:

\begin{enumerate}
\item \textit{Positivity.} The angular Sturm-Liouville operator requires 
$\nu(\nu+1) \geq 0$, satisfied only for $\nu \geq 0$ or $\nu \leq -1$. The 
interior region $-1 < \nu < 0$ is forbidden.

\item \textit{Vanishing fields.} All electromagnetic field components vanish 
identically at $\nu = -1$. The factor $\nu(\nu+1) \to 0$ annihilates the radial 
component, while the constant angular function $P_{-1}(\cos\theta) = 1$ has zero 
derivative, annihilating the tangential components.

\item \textit{Monopole-type singularity.} The $\nu = -1$ configuration connects 
formally to electrostatic monopole structures, which are excluded from the cavity 
mode spectrum by energy considerations.
\end{enumerate}

Nevertheless, while the electromagnetic field vanishes at $\nu = -1$, the Debye 
potential $\Pi = \cos(kr)/kr$ remains a well-defined, non-trivial solution to the 
scalar Helmholtz equation with a $1/r$ singularity at the origin. The vanishing of 
fields reflects the kernel structure of the curl-curl operator for spherically 
symmetric configurations, not a triviality of the potential itself. This distinction 
between potential and field connects to broader questions in gauge theory and cavity 
quantization that remain subjects of active discussion.

The analysis delineates the boundary between propagating electromagnetic modes and 
static field configurations in spherical geometry. For practical applications 
involving continuous-$\nu$ modes accessed through conical boundaries, the physical 
spectrum remains bounded by $\nu \geq 0$, with the physical spectrum bounded below by $\nu = m = 0$. However, 
$\nu = 0$ itself yields no electromagnetic field; the first physical 
zonal mode on the full sphere occurs at $\nu = 1$.

\end{document}